\documentclass[showpacs,aps,graphicx]{revtex4}
\usepackage{graphicx}
\usepackage{amsmath}
\usepackage{amssymb}
\usepackage[titletoc]{appendix}

\begin{document}

\title{Decomposition of orthogonal matrix and synthesis of two-qubit and three-qubit orthogonal gates}

\author{Hai-Rui Wei,$^{1,2}$ Yao-Min Di$^{1,}$\footnote{ Email address: yaomindi@sina.com}}

\address{ $^1$School of Physics $\&$ Electronic Engineering, Xuzhou Normal
University, Xuzhou 221116,  China \\
$^2$Department of Physics, Beijing Normal University, Beijing
100875, China}


\begin{abstract}
The decomposition of matrices associated to two-qubit and
three-qubit orthogonal gates is studied, and based on the
decomposition the synthesis of these gates is investigated. The
optimal synthesis of general two-qubit orthogonal gate is obtained.
For two-qubit unimodular orthogonal gate, it requires at most 2 CNOT
gates and 6 one-qubit $R_y$ gates. For the general three-qubit
unimodular orthogonal gate, it can be synthesized by 16 CNOT gates
and 36 one-qubit $R_y$ and $R_z$ gates
 in the worst case.\\
\end{abstract}


\maketitle

\section{Introduction}\label{sec1}
In quantum computing, the algorithms are commonly described by the
quantum circuit model \cite{1}. The building blocks of quantum
circuits are quantum gates, i.e., unitary transformations acting on
a set of qubits. In 1995, Barenco et al showed that any qubit
quantum circuit can be decomposed into a sequence of one-qubit gates
and CNOT gates \cite{2}. The process of constructing quantum
circuits by these elementary gates is called synthesis by some
authors. The complexity of quantum circuit can be measured in terms
of the number of CNOT and one-qubit elementary gates required.
Achieving gate arrays of less complexity is crucial not only because
it reduces resource, but it also reduces errors.

Decomposition of matrix plays very important role to synthesize and
optimize quantum gates. Based on Cartan decomposition \cite{3,4},
the synthesis, optimization and ``small circuit'' structure of
two-qubit gate are well solved \cite{5,6,7,8,9}. To implement the
general two-qubit gate, it requires at most 3 CNOT gates and 15
elementary one-qubit gates from the family
$\left\{R_{y},R_{z}\right\}$ \cite{7,8}.

Unfortunately, the aforementioned optimal synthesis of most general
two-qubit quantum gates have not yet led to similarly tight results
for three-qubit gates. Based on one of Cartan decompositions for
multi-qubit system, Khaneja-Glaser decomposition (KGD) \cite{4},
Vatan and Williams get the result that a general three-qubit quantum
gate can be synthesized using at most 40 CNOT gates and 98 one-qubit
$R_y$ and $R_z$ gates \cite{10}. Using the modified KGD, the results
have been improved in \cite{11}, that is it requires at most 26 CNOT
gates and 73 one-qubit $R_y$ and $R_z$ gates. Now the best known
result is based on quantum Shannon decomposition (QSD) \cite{12}
proposed by Shende, Bullock and Markov, it requires at most 20 CNOT
gates. According to the result of multi-qubit case, the best known
theoretical lower bound on CNOT gate cost for general three-qubit
gates is 14 \cite{8}. However, no circuit construction yielding
these numbers of CNOT gates has been presented in the literature.

The orthogonal gate is an important class of gate, the matrix
corresponding to the gate is orthogonal. For example, classical
reversible logic circuits have a long history \cite{13} and are a
necessary subclass whose realization is required for any quantum
computer to be universal. The matrix elements of them are all real,
so they are orthogonal. Utilizing the basic property of magic basis,
in 2004, Vatan and Williams investigated the synthesis of two-qubit
orthogonal gate in \cite{7}. The result is that the synthesis of the
unimodular orthogonal gate requires at most 2 CNOT gates and 12
elementary one-qubit gates. As for the non-unimodular orthogonal
gate, that is its matrix determinant is equal to minus one, it
requires at most 3 CNOT gates and 12 elementary one-qubit gates
\cite{7}. The number of the one-qubit gates required can still be
reduced further. Moreover, no articles discuss the synthesis of
general orthogonal three-qubit quantum gates yet.

In this work, we devote to investigating the synthesis of two-qubit
and three-qubit orthogonal gates. For this purpose, we study the
Cartan decomposition of matrix for these gates first. Based on the
particular decompositions, the two kinds of synthesis are obtained.
For two-qubit unimodular orthogonal gate, it requires at most 2 CNOT
gates and 6 one-qubit $R_{y}$  gates, beating an earlier bound of 2
CNOT gates and 12 one-qubit elementary gates. The numbers required
for one-qubit gate and CNOT gate are all reach the lower bound. For
three-qubit unimodular orthogonal gate, it can be synthesized by 16
CNOT gates and 36 one-qubit $R_y$ and $R_z$ gates in the worst case.

This paper is organized as follows. The concept of Cartan
decomposition and its application in quantum information science
(QIS) are briefly introduced in Section \ref{sec2}. Based on a kind
of Cartan decomposition of special orthogonal group
$\mathrm{SO}(4)$, we provide an optimal synthesis of general
two-qubit orthogonal gate in Section \ref{sec3}. The decomposition
of the $\mathrm{SO(8)}$ group associated to three-qubit unimodular
orthogonal gate is investigated in Section \ref{sec4}. The synthesis
of general three-qubit unimodular orthogonal gate is studied in
Section \ref{sec5}. It is first time to discuss the synthesis of
this kind gate. A brief conclusion is made in Section \ref{sec6}.

\section{Cartan Decomposition and Its Application in QIS}\label{sec2}
\noindent The Cartan decomposition of Lie group \cite{3} depends on
the decomposition of its Lie algebra. A Cartan decomposition of a
real semisimple Lie algebra $\mathfrak{g}$ is the decomposition
\begin{equation}                                        \label{eq.1}
\mathfrak{g}=\mathfrak{l}\oplus \mathfrak{p},
\end{equation}
where  $\mathfrak{p}$ is the orthogonal complement of $\mathfrak{l}$
with respect to the Killing form, $\mathfrak{l}$ and $\mathfrak{p}$
satisfy the commutation relations:
\begin{equation}                                             \label{eq.2}
[\mathfrak{l},\mathfrak{l}]\subseteq \mathfrak{l},
[\mathfrak{l},\mathfrak{p}]\subseteq \mathfrak{p},
[\mathfrak{p},\mathfrak{p}]\subseteq \mathfrak{l}.
\end{equation}
$\mathfrak{l}$ is a Lie subalgebra of $\mathfrak{g}$. A maximal
Abelian subalgebra contained in $\mathfrak{p}$ is called a Cartan
subalgebra of the pair $(\mathfrak{g},\mathfrak{l})$ denoted as
$\mathfrak{a}$. Then using the relation between Lie group and Lie
algebra, every element  $X$ of the Lie group $G$ can be written as
\begin{equation}
X=K_{1}AK_{2} ,                         \label{eq.3}
\end{equation}
where  $G=e^\mathfrak{g}$, $K_{1}$, $K_{2}\in e^\mathfrak{l} $ and
$A\in e^\mathfrak{a}$.

There are many kinds of Cartan decomposition for semisimple Lie
groups. Now the main application in quantum information science is
the decomposition of $\mathrm{SU}(2^{n})$ group for multi-qubit
system, i.e. Khaneja-Glaser Decomposition (KGD) \cite{4}. Moreover
there are some other decompositions, such as Concurrence Canonical
Decomposition (CCD) \cite{14,15} which is a decomposition of
$\mathrm{SU}(2^{n})$ group too, the Odd-Even Decomposition (OED)
\cite{16}, which is a generalization of CCD to more general
multipartite quantum system case. Some kinds of Cartan decomposition
for a bipartite high dimension quantum system were discussed in
\cite{17,18,19}. These Cartan decompositions have been applied in
the synthesis and implementation of quantum logic gates
\cite{10,11,20,21}, the entanglement of multipartite quantum systems
\cite{14,15}, etc. But we need to find new suitable algebraic
structures of Cartan decomposition to meet the purpose here.

\section{Optimal Synthesis of General Two-Qubit Orthogonal Gates}\label{sec3}
\noindent We now consider the decomposition of 4 dimensional special
orthogonal group $\mathrm{SO}(4)$ associated to the two-qubit
unimodular orthogonal gate. Difference from that in \cite{7}, the
$\mathrm{so}(4)$ Lie algebra is constructed as
\begin{eqnarray}                                                                                         \label{eq.4}
\mathrm{so}(4):=span\{iI\otimes \sigma_{y},i\sigma_{y}\otimes
I,i\sigma_{x}\otimes \sigma_{y}, i\sigma_{y}\otimes \sigma_{x},
       i\sigma_{z}\otimes \sigma_{y},i\sigma_{y}\otimes \sigma_{z}\},
 \end{eqnarray}
in which each basis vector involves a $\sigma_{y}$ matrix. A kind of
Cartan decomposition of $\mathrm{so}(4)$ algebra is that
\begin{eqnarray}                                     \label{eq.5}
\mathrm{so}(4)=\mathfrak{l}\oplus \mathfrak{p},
\end{eqnarray}
with
\begin{eqnarray}                                                \label{eq.6}
\mathfrak{l}:=span\left\{iI\otimes \sigma_{y},i\sigma_{y}\otimes
I\right\},
\end{eqnarray}
\begin{eqnarray}                                                          \label{eq.7}
\mathfrak{p}:=span\left\{i\sigma_{x}\otimes
\sigma_{y},i\sigma_{y}\otimes \sigma_{x},
   i\sigma_{z}\otimes \sigma_{y},i\sigma_{y}\otimes \sigma_{z}\right\}.
\end{eqnarray}
where $\mathfrak{l}$ is a Lie subalgebra and
$\mathfrak{p}=\mathfrak{l}^{\perp }$. Its Cartan subalgebra is
\begin{eqnarray}                                                           \label{eq.8}
\mathfrak{a}:=span\left\{i\sigma_{x}\otimes
\sigma_{y},i\sigma_{y}\otimes
       \sigma_{z}\right\}.
\end{eqnarray}
Utilizing the relation between Lie group and Lie algebra, the Cartan
decomposition of Lie group $\mathrm{SO}(4)$ can be obtained. For
every element $X \in \mathrm{SO}(4)$, we have
 \begin{equation}                                           \label{eq.9}
X=K_{1}AK_{2},
\end{equation}
where $K_{1},K_{2}\in \mathrm{SO}(2) \otimes \mathrm{SO}(2)$, and
$A$ is a two-qubit operation of the form
 \begin{eqnarray}                                           \label{eq.10}
A(a,b)=\exp\left(-i\left(a\sigma_{x}\otimes \sigma_{y}+b
\sigma_{y}\otimes \sigma_{z}\right)\right),
 \end{eqnarray}
where $a,b \in R $.

The $A(a,b)$ can be represented by the synthesis of elementary gates
as
\begin{eqnarray}                                                  \label{eq.11}
A=C_2^1\cdot R_{y}^{(1)}(b)\cdot R_{y}^{(2)}(a)\cdot C_2^1.
 \end{eqnarray}

\begin{figure}[!h]
\begin{center}
\includegraphics[width=5.0 cm,angle=0]{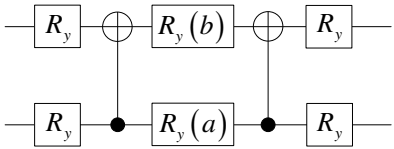}
\caption{A circuit for computing the general two-qubit unimodular
orthogonal gate.} \label{Figure1}
\end{center}
\end{figure}

Here and afterwards $C_{j}^{i}$ denotes the CNOT gate that control
on the $j$-th qubit and target on the $i$-th qubit, and
$R_{\alpha}^{(i)}(a)=\exp\left(-i a \sigma_{\alpha}\right)$
$\left(\alpha \in x,y,z\right)$ is an elementary one-qubit gate
acting on the $i$-th qubit. Combing Eqs.(9), (10) and (11), we can
get the synthesis of general two-qubit orthogonal gate as in
Fig.\ref{Figure1}, it requires at most 2 CNOT gates and 6 one-qubit
$R_{y}$ gates. As for the non-unimodular orthogonal gate (the
determinant is equal to minus one), it requires at most 3 CNOT gates
and 6 one-qubit $R_{y}$ gates. The 2 CNOT gates is optimal for CNOT
gate cost of two-qubit orthogonal gate, and it has been proved in
\cite{9}. Since a $\mathrm{SO}(4)$ matrix has 6 independent
parameters, it needs at least 6 elementary one-qubit gates to load
them. So the synthesis of general two-qubit orthogonal gate here is
optimal both for CNOT gates and elementary one-qubit gates.

\section{ Decomposition of General Three-qubit Unimodular Orthogonal Gate}\label{sec4}
\noindent The matrices of any general three-qubit unimodular
orthogonal gate are elements of special orthogonal group
$\mathrm{SO}(8)$. We construct the $\mathrm{so}(8)$ Lie algebra
first. Taking \textbf{AI} type of Cartan decomposition \cite{3} on
$\mathrm{u}(4)$ and $\mathrm{u}(2)$ Lie algebra,
\begin{eqnarray}                                            \label{eq.12}
\mathrm{u}(4)=i \sigma^{(1)} \oplus  i S^{(1)},
\end{eqnarray}
\begin{eqnarray}                                             \label{eq.13}
\mathrm{u}(2)=i \sigma^{(2)} \oplus i S^{(2)},
 \end{eqnarray}
with
\begin{eqnarray}                                                                                    \label{eq.14}
 i\sigma^{(1)}:=span \{i I\otimes \sigma_{x},iI\otimes\sigma_{y},i I \otimes \sigma_{z},
                i\sigma_{x} \otimes I, i \sigma_{y} \otimes I,i \sigma_{z} \otimes I \},
\end{eqnarray}
\begin{eqnarray}                                                                                      \label{eq.15}
i S^{(1)}:=span\left\{i\sigma_{x,y,z}\otimes \sigma_{x,y,z}, i I
\right\},
 \end{eqnarray}
\begin{eqnarray}                                            \label{eq.16}
i\sigma^{(2)}:=span\left\{i\sigma_{y}\right\},\
iS^{(2)}:=span\{i\sigma_{x},i\sigma_{z},i I\}.
\end{eqnarray}
 A set of basis for a Lie algebra is given by 28 tensor products of the form
\begin{eqnarray}                                                  \label{eq.17}
F:=i\sigma^{(1)}\otimes S^{(2)} and \  i S^{(1)}\otimes
\sigma^{(2)}.
 \end{eqnarray}
Using the transformation matrix in \cite{22,23}:
\begin{eqnarray}                                      \label{eq.18}
\mathcal {M}=\frac{1}{\sqrt{2}}\left(\begin{array}{cccc}
1&i&0&0\\
0&0&i&1\\
0&0&i&-1\\
1&-i&0&0\\
\end{array}\right)\otimes I_{2},
\end{eqnarray}
the $\mathrm{so}(8)$ Lie algebra can be obtained by $\mathcal
{M}^{\dag } \cdot span\left\{F\right\} \cdot \mathcal{M} $. So the
Lie algebra $\mathfrak{g}:=span\left\{F\right\}$ is isomorphic to
$\mathrm{so}(8)$, and the basis in Eq.(\ref{eq.18}) can be called as
magic basis of $\mathrm{so}(8)$ algebra.

Then we take Cartan decomposition of the Lie algebra $\mathfrak{g}$
as Eq.(\ref{eq.1}), with
 \begin{eqnarray}                                             \label{eq.19}
\mathfrak{l}:=span\{iI\otimes \sigma_{x,y,z}\otimes I,
       i\sigma_{x,y,z}\otimes I \otimes I,
       i I \otimes \sigma_{x,y,z}\otimes \sigma_{z},
       i \sigma_{x,y,z} \otimes I \otimes \sigma_{z}\},
\end{eqnarray}
\begin{eqnarray}                                                      \label{eq.20}
\mathfrak{p}:=span\{iI\otimes \sigma_{x,y,z}\otimes
    \sigma_{x},i\sigma_{x,y,z}\otimes I \otimes \sigma_{x},
    i I \otimes I\otimes \sigma_{y},
    i \sigma_{x,y,z} \otimes \sigma_{x,y,z} \otimes
    \sigma_{y}\}.
 \end{eqnarray}
The $\mathfrak{l}$ is isomorphic to $\mathrm{so}(4) \oplus
\mathrm{so}(4)$. The Cartan subalgebra of the pair
$(\mathfrak{g},\mathfrak{l})$ can be chosen as
 \begin{eqnarray}                                                      \label{eq.21}
\mathfrak{a}:= span\{i\sigma_{x}\otimes \sigma_{x} \otimes
\sigma_{y},
   i\sigma_{y}\otimes \sigma_{y} \otimes \sigma_{y},
   i\sigma_{z}\otimes \sigma_{z} \otimes \sigma_{y},
   iI\otimes I \otimes \sigma_{y} \}.
 \end{eqnarray}
Using the formula
\begin{equation}                                                     \label{eq.22}
\left[A \otimes B,C \otimes D\right]=\frac{1}{2}\left(\{A,C\}
\otimes [B,D]+[A,C]\otimes\{B,D\}\right),
\end{equation}
it is easy to verify that the $\mathfrak{l}$ and $\mathfrak{p}$ in
Eqs.(\ref{eq.19}, \ref{eq.20}) satisfy the conditions of the Cartan
decomposition in Eq.(\ref{eq.2}).

Lie subalgebra $\mathfrak{l}$ could be decomposed further
\begin{eqnarray}                                              \label{eq.23}
\mathfrak{l}=\mathfrak{l}^{(1)}\oplus \mathfrak{p}^{(1)},
 \end{eqnarray}
with
 \begin{eqnarray}                                            \label{eq.24}
\mathfrak{l}^{(1)}:=span\left\{iI\otimes \sigma_{x,y,z}\otimes I,
      i\sigma_{x,y,z}\otimes I \otimes I\right\},
 \end{eqnarray}
 \begin{eqnarray}                                                 \label{eq.25}
\mathfrak{p}^{(1)}:=span\left\{i I \otimes \sigma_{x,y,z}\otimes
\sigma_{z},
     i\sigma_{x,y,z} \otimes I \otimes \sigma_{z} \right\}.
 \end{eqnarray}
The $\mathfrak{l}^{(1)}$ is isomorphic to $\mathrm{so}(4)$. Its
Cartan subalgebra can be chosen as
 \begin{eqnarray}                                           \label{eq.26}
\mathfrak{a}^{(1)}:=span\left\{i I \otimes \sigma_{z} \otimes
\sigma_{z}, i \sigma_{z} \otimes I \otimes \sigma_{z}\right\}.
\end{eqnarray}
From the correspondence between Lie group and Lie algebra and the
conjugative transformation, we get the Cartan decomposition of Lie
group $\mathrm{SO}(8)$: any element of the group can be decomposed
as
 \begin{eqnarray}                                            \label{eq.27}
X=\mathcal {M}^{\dag} K_{1} A_{1}^{(1)} K_{2} A K_{3} A_{2}^{(1)}
K_{4} \mathcal {M}.
\end{eqnarray}
Here $K_{i}\in \mathrm{SU}(2)\otimes \mathrm{SU}(2)$, $A$ and
$A_{i}^{(1)}$ are the Abelian subgroup associated to Cartan
subalgebra
 $\mathfrak{a}$ and  $\mathfrak{a}^{(1)}$ respectively.

\section{ Synthesis of General Three-qubit Unimodular Orthogonal Gate}\label{sec5}
\begin{figure}[!h]
\begin{center}
\includegraphics[width=8.0 cm,angle=0]{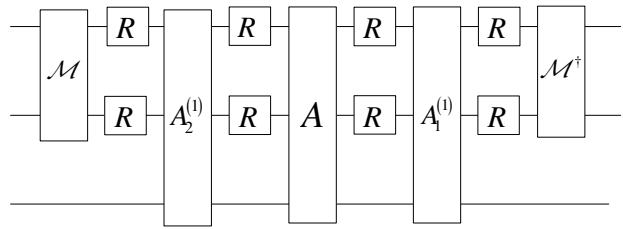}
\caption{Decomposition of the general three-qubit unimodular
orthogonal gate, $R=R_{z}R_{y}R_{z}$.}\label{Figure2}
\end{center}
\end{figure}

\noindent Based on the discussion in Section \ref{sec4}, the
decomposition of general three-qubit orthogonal gate is shown in
Fig.\ref{Figure2}, where $R=R_{z}(\theta)R_{y}(\varphi)R_{z}(\psi)
\in \mathrm{SU}(2)$ and

\begin{figure}[!h]
\begin{center}
\includegraphics[width=4.0 cm,angle=0]{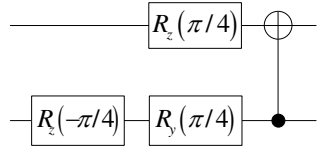}
\caption{A circuit for computing the magic matrix $\mathcal{M}$.}
\label{Figure3}
\end{center}
\end{figure}

\begin{eqnarray}                                               \label{eq.28}
A(a,b,c,d)=\exp\{-i(a\sigma_{x}\otimes
                 \sigma_{x}\otimes\sigma_{y}+
                 b \sigma_{y}\otimes \sigma_{y}\otimes
                 \sigma_{y}\
                 + c \sigma_{z}\otimes \sigma_{z}\otimes \sigma_{y}+
                 d I\otimes I\otimes \sigma_{y})\},
\end{eqnarray}
\begin{eqnarray}                                                             \label{eq.29}
A^{(1)}(\alpha,\beta)=\exp\left\{-i(\alpha I\otimes
\sigma_{z}\otimes \sigma_{z}+ \beta \sigma_{z}\otimes I \otimes
\sigma_{z}\right\}.
\end{eqnarray}

\begin{figure}[!h]
\begin{center}
\includegraphics[width=11.0 cm,angle=0]{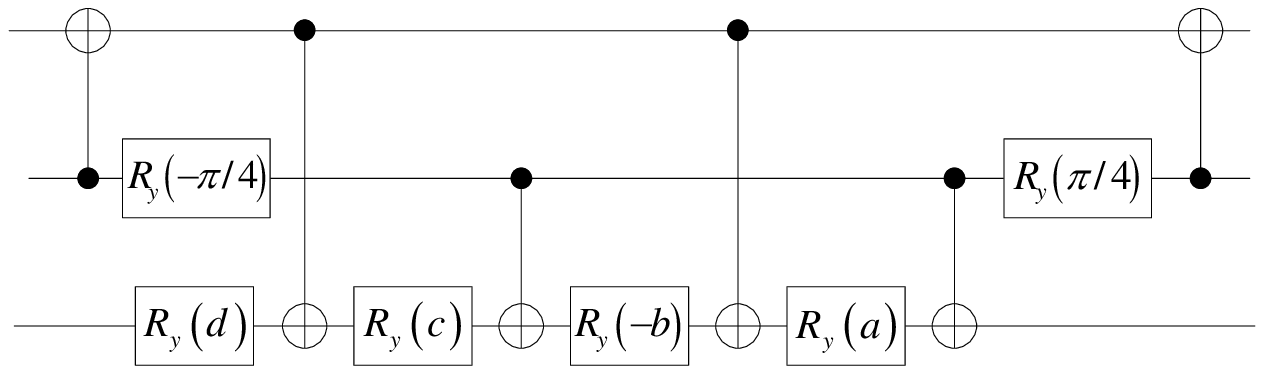}
\caption{A circuit for computing the unitary operation $A$.}
\label{Figure4}
\end{center}
\end{figure}

\begin{figure}[!h]
\begin{center}
\includegraphics[width=5.0 cm,angle=0]{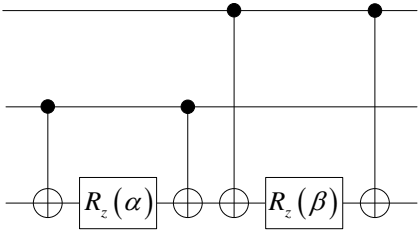}
\caption{A circuit for computing the unitary operation $A^{(1)}$.}
\label{Figure5}
\end{center}
\end{figure}

The synthesis of transformation matrix $\mathcal{M}$ is given in
\cite{7} shown in Fig.\ref{Figure3}, that is
\begin{eqnarray}                                                       \label{eq.30}
\mathcal {M}=C_{2}^{1}\cdot R_{z}^{(1)}\left(\frac{\pi}{4}\right)
           \cdot R_{y}^{(2)}\left(\frac{\pi}{4}\right)
           \cdot R_{z}^{(2)}\left(-\frac{\pi}{4}\right).
\end{eqnarray}
The $A$ can be expressed as
\begin{eqnarray}                                                  \label{eq.31}
A(a,b,c,d)=\mathcal {M} \cdot\tilde{A}(a,b,c)\cdot \mathcal
{M}^{\dag}\cdot R_{y}^{(3)}(d),
\end{eqnarray}
here
\begin{center}
\begin{eqnarray}                                                   \label{eq.32}
\tilde{A}(a,b,c)&=&\exp\{-i(a I \otimes \sigma_{z}\otimes
\sigma_{y}-
                 b \sigma_{z}\otimes \sigma_{z}\otimes \sigma_{y}+
                 c \sigma_{z}\otimes I \otimes \sigma_{y})\}\nonumber  \\
 &=&\exp\left \{-i
\left(\begin{array}{cccc}
a-b+c&0&0&0\\
0&b-a+c&0&0\\
0&0&a+b-c&0\\
0&0&0&-a-b-c\\
\end{array}
\right)\otimes \sigma_{y}\right\}.
\end{eqnarray}
\end{center}
Since the Cartan subalgebra is commutative, we can break down the
synthesis of $\tilde{A}(a,b,c)$ into the following operations:
\begin{eqnarray}                                             \label{eq.33}
 \tilde{A}_{1}(a)=\exp\{-i a I \otimes
                   \sigma_{z}\otimes \sigma_{y}\},
\end{eqnarray}
\begin{eqnarray}                                             \label{eq.34}
 \tilde{A}_{2}(-b)=\exp\{i b \sigma_{z}\otimes
                    \sigma_{z}\otimes\sigma_{y}\},
\end{eqnarray}
\begin{eqnarray}                                             \label{eq.35}
 \tilde{A}_{3}(c)=\exp\{-i c \sigma_{z}\otimes I
                   \otimes\sigma_{y}\}.
\end{eqnarray}
And we have
\begin{eqnarray}                                             \label{eq.36}
\tilde{A}_{1}(a)=C_{2}^{3}\cdot R_{y}^{(3)}(a)\cdot C_{2}^{3},
\end{eqnarray}
\begin{eqnarray}                                             \label{eq.37}
\tilde{A}_{2}(-b)= C_{1}^{3}  \cdot C_{2}^{3} \cdot
R_{y}^{(3)}(-b)\cdot C_{2}^{3}\cdot
 C_{1}^{3},
\end{eqnarray}
\begin{eqnarray}                                                \label{eq.38}
\tilde{A}_{3}(c)= C_{1}^{3}  \cdot  R_{y}^{(3)}(c)\cdot C_{1}^{3}.
\end{eqnarray}
By putting Eqs.(\ref{eq.36}), (\ref{eq.37}) and (\ref{eq.38})
together, we get
\begin{eqnarray}                                             \label{eq.39}
\tilde{A}(a,b,c)=C_{2}^{3}\cdot R_{y}^{(3)}(a)\cdot C_{1}^{3}\cdot
R_{y}^{(3)}(-b)\cdot
 C_{2}^{3}\cdot
R_{y}^{(3)}(c)\cdot C_{1}^{3}.
\end{eqnarray}
Here the identity $C_{1}^{3} \cdot C_{2}^{3}=C_{2}^{3} \cdot
C_{1}^{3}$ is used. Combining Eqs.(\ref{eq.30}), (\ref{eq.31}) and
(\ref{eq.39}), we have
\begin{eqnarray}                                       \label{eq.40}
A(a,b,c,d)&=&C_{2}^{1}\cdot
               R_{y}^{(2)}\left(\frac{\pi}{4}\right)\cdot
               C_{2}^{3}\cdot
               R_{y}^{(3)}\left(a\right)\cdot
               C_{1}^{3}\cdot
               R_{y}^{(3)}\left(-b\right)\cdot \nonumber  \\&&
               C_{2}^{3} \cdot
               R_{y}^{(3)}\left(c\right)\cdot
               C_{1}^{3}\cdot
               R_{y}^{(2)}\left(-\frac{\pi}{4}\right)\cdot
               C_{2}^{1}\cdot
               R_{y}^{(3)}(d).
\end{eqnarray}
and its circuit shown in Fig.\ref{Figure4}. Since $R_{z}$ gates
commute with the control qubit of the CNOT gate, here the $R_{z}$
gates in $\mathcal {M}$ and $\mathcal{M}^{\dag}$ are canceled. The
synthesis of $A^{(1)}$ is shown in Fig.\ref{Figure5}, that is
\begin{eqnarray}                                                \label{eq.41}
A^{(1)}(\alpha,\beta)=C_{1}^{3}\cdot R_{z}^{(3)}(\beta)\cdot
C_{1}^{3}\cdot C_{2}^{3}\cdot R_{z}^{(3)}(\alpha)\cdot C_{2}^{3}.
\end{eqnarray}

Putting all these pieces together, we get that 16 CNOT gates and 36
one-qubit $R_{y}$ and $R_{z}$ gates at most are sufficient to
synthesize general three-qubit orthogonal gate. For the same reason
mentioned above, the two $R_{z}$ gates acting on same qubit
neighbored $A^{(2)}$ have been combined to one. Here, each
$\mathcal{M}$ requires 1 CNOT gates and 3 one-qubit $R_y$ and $R_z$
gates, $A$ requires 6 CNOT gates and 6 one-qubit $R_y$ gates, each
$A^{(1)}$ requires 4 CNOT gates and 2 one-qubit $R_y$ and $R_z$
gates, and 8 $R$ gates require 20 one-qubit $R_y$ and $R_z$ gates.

\section{Conclusions}\label{sec6}
\noindent Based on the decomposition of matrices, the synthesis of
two-qubit and three-qubit orthogonal gates is investigated. For
two-qubit orthogonal gate, we get optimal result, which requires at
most 2 CNOT gates and 6 one-qubit $R_{y}$ gates, beating an earlier
bound of 2 CNOT gates and 12 one-qubit gates. For the three-qubit
unimodular orthogonal gate, it requires 16 CNOT gates and 36
one-qubit gates from the family $\left\{R_{y},R_{z}\right\}$  in the
worst case. There are abundant algebraic structures for matrix
decomposition of three-qubit orthogonal gate. We have many ways to
investigate the synthesis of general three-qubit orthogonal gate.
The result given here is the best one we have got, although we can
not affirm that is optimal yet. The synthesis of general three-qubit
gate has been studied in some literatures \cite{10,11,12}, the
orthogonal gate is an important class of gate of them. So the work
here is essentially on the ``small circuit'' issue of three-qubit
gates, which is first investigated in this paper. Different from
two-qubit gate, how to get optimal quantum circuit for general
three-qubit gate has not been well solved and is worthy studying
further.

\section*{ACKNOWLEDGEMENTS} \noindent The work was supported by the
Project of Natural Science Foundation of Jiangsu Education Bureau,
China(Grant No.09KJB140010).

\end{document}